\documentclass[runningheads]{llncs}
\usepackage{graphicx}
\usepackage{todonotes}
\usepackage{soul}
\begin{document}

\newcommand{\dragos}[1]{\textcolor{red}{#1}}
\newcommand{\ziyad}[1]{\textcolor{blue}{#1}}
\newcommand{\manuel}[1][1]{\textcolor{red}{#1}}
\newcommand{\rasheed}[1][1]{\textcolor{magenta}{#1}}

%
%\title{Using machine learning for generating value to software
%development: A survey of the state-of-the-art}
% \title{Software Design of an Open-source \\ Framework for EAV Charging}
%
\title{A Machine to Machine framework for the charging of Electric Autonomous Vehicles}
\titlerunning{Inno-EAV}
% If the paper title is too long for the running head, you can set
% an abbreviated paper title here
%

\author{
Ziyad Elbanna
\and Ilya Afanasyev
\and Luiz J.P. Ara\'ujo
\and Rasheed Hussain
\and Mansur Khazeev
\and Joseph Lamptey
\and Manuel Mazzara 
\and Swati Megha
\and Diksha Moolchandani 
\and Dragos Strugar
}

\authorrunning{Ziyad Elbanna et al.}

% First names are abbreviated in the running head.
% If there are more than two authors, 'et al.' is used.
%
\institute{Innopolis University, Innopolis 420500, Russia\\
}
\maketitle              % typeset the header of the contribution
\vspace{-5 mm}
\begin{abstract}
Electric Autonomous Vehicles (EAVs) have gained increasing attention of industry, governments and scientific communities concerned about issues related to classic transportation including accidents and casualties, gas emissions and air pollution, intensive traffic and city viability. One of the aspects, however, that prevent a broader adoption of this technology is the need for human interference to charge EAVs, which is still mostly manual and time-consuming. This study approaches such a problem by introducing the Inno-EAV, an open-source charging framework for EAVs that employs machine-to-machine (M2M) distributed communication. The idea behind M2M is to have networked devices that can interact, exchange information and perform actions without any manual assistance of humans. The advantages of the Inno-EAV include the automation of charging processes and the collection of relevant data that can support better decision making in the spheres of energy distribution. In this paper, we present the software design of the framework, the development process, the emphasis on the distributed architecture and the networked communication, and we discuss the back-end database that is used to store information about car owners, cars, and charging stations.
\end{abstract}


%\clearpage
\section{Introduction}

Electric Autonomous Vehicles (EAV) have been ubiquitously replacing standard vehicles since their invention in the 20\textsuperscript{th} century \cite{history}. EAV technology is made possible by the synergy of Vehicular Communication \cite{Ting2009}, Intelligent Transportation Systems (ITS) and Big Data Analytics \cite{Zhu2019}. Vehicular Communication systems are computer networks in which both vehicles and roadside units represent communicating nodes exchanging information about safety and traffic. ITSs are applications providing innovative services for traffic management and informed decisions. Research has shown that these technologies represent a practical approach to solving such transportation issues as accidents \cite{nhtsa2018reasons}, traffic congestion, and harmful gas emissions \cite{KYRIAKIDIS2015127,m2mwebsite}. 

Autonomous cars are already pervading our roads in the form of pilot execution from different stakeholders and are speculated to be on the verge of commercialisation \cite{Hussain2019}. It is also worth mentioning that autonomous cars are poised to have profound social and economic impacts on our lives, as well as a positive impact on the environment \cite{Hussain2018}.  The EAV technology is now emerging as an enabler to establish the foundation for the Machine-to-Machine (M2M) economy, i.e. networked devices that can interact, exchange information and perform actions without any manual assistance of humans. For this to happen, the market should be able to offer appropriate charging services without involving humans \cite{StrugarHMRAL19,StrugarHMRLM18,hua2018,afanasyev2019towards}.

%An order of magnitude or big gains in software engineering process has been a perennial concern. Software development lifecycles and orchestration has been some of the gains in developing a quality software product fast. According to Fred Brooks, continuous integration and delivery might be the silver bullet that enhances the delivery of better, cheaper, high quality features, and complete software to the end user \cite{ozkaya2019devops}. He further asserts that the place of automation, DevOps, is well-bounded and repetitive tasks.

%Meanwhile, Bass et. al. \cite{bass2015devops} defines DevOps as: "a set of practices intended to reduce the time between committing a change to a system and the change being placed into normal production, while ensuring high quality". This optimisation and delivery of value to the end user is achieved through the support of tools and infrastructure \cite{bass2015devops} such as automated code commits to a repository and automated testing on servers or AWS lambda functions (server-less).

In this paper, we investigate the area of Intelligent Vehicle and Transportation Systems and present the design of a versatile open-source framework developed for improving and automating the charging process of electric vehicles and establishing the foundation of M2M economy using WiFi networks. Inno-EAV is available as a set of open-source modules which can be used independently or concomitantly  by DevOps teams.  It offers a repository that can be cloned  into a project for continuous integration and development \cite{InnoEAV2019}.

\section{The Inno-EAV Architectural Layers}

%In the recent years, wireless M2M communication \cite{williams2017,afanasyev2019} has extended the features of wired communication machines to include onboard Global Positioning System (GPS), flexible surface mounting of land grid array, embedded M2M optimised smart cards (e.g. phone Subscriber Identity Modules (SIMs) known as M2M Identification Modules (MIMs) \cite{moller2019connected}. To enable such a vast number of possible applications, IoT application developers often gravitate toward the embedded Java as their programming language of choice due to its portability, flexibility and versatility. M2M systems often use public networks and access methods (e.g. cellular or Ethernet) to make it more cost-effective. The main components of M2M systems include sensors, Radio Frequency Identification (RFID), Wireless-Fidelity (WiFi) or cellular communications link and autonomic computing software programmed to aid the network device to interpret data and make decisions. Such M2M applications translate the data, which can trigger pre-programmed automated actions.

\begin{figure}[ht]
    \centering
    \includegraphics[width=1.0\textwidth]{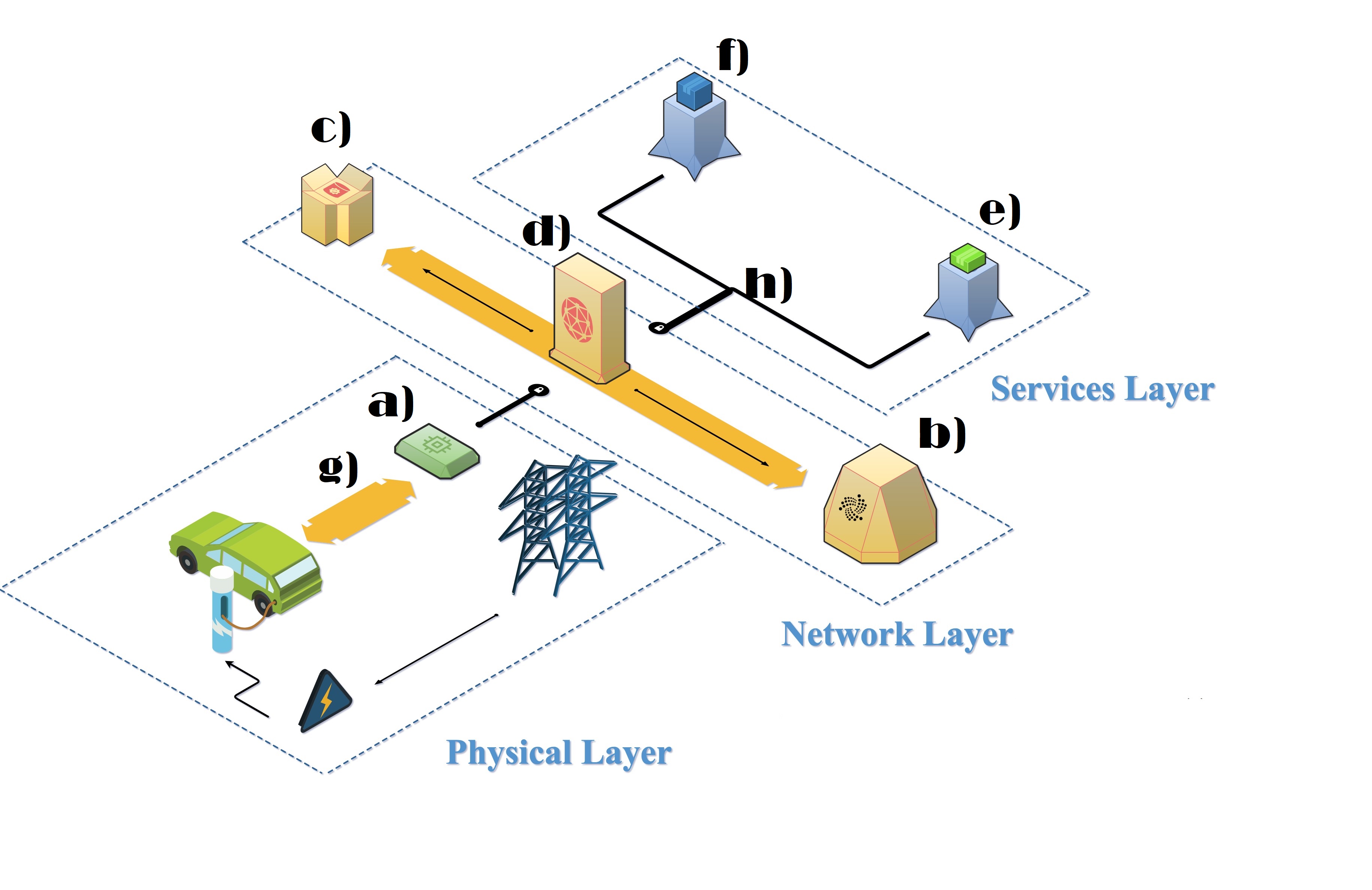}
    \caption{\textbf{Inno-EAV Architecture.} Currently Inno-EAV consists of 3 layers dedicated to high performance as explained and depicted in \cite{StrugarHMRAL19}, with labels refering to: a)Main Controller b)IOTA's tangle, c)Local database, d)Bidirectional Communication, e)EAV's app, f)SP application, g) Component connection in one layer and h) Secured connection between two layers respectively. }
    \label{fig:System Architecture}
\end{figure}

The Inno-EAV framework is constituted of three layers (Fig. \ref{fig:System Architecture}):

\begin{itemize}
\item \textbf{Physical layer}. It encapsulates all the hardware components embedded in the charging station (CS) used for sensing and gathering information about the charging process, as well as communicating with external entities.
\item \textbf{Network Layer}. Since our scenario is composed of autonomous, dynamic decision-making CSs and vehicles, we suggest the communication between the entities in the system be entirely peer-to-peer (P2P) with distributed ledger technologies (DLTs) as a possible solution to store both data and value transactions \cite{Bencic18,Burkhardt18}. We also consider Message Queuing Telemetry Transport (MQTT) connectivity protocol to be the most appropriate choice for inter-device communication. In this way, the network layer would be able to transfer sensor data from the Physical Layer to the Services Layer rapidly and securely \cite{StrugarHMRAL19,StrugarHMRLM18}.
\item \textbf{Services Layer}. On top of the architecture sits the Services Layer, which makes use of the two layers described above to deliver services to both consumers and service providers. Such services include:

\begin{itemize}
   \item Charging services for EAVs;
   \item Data Analysis for Service Providers.
\end{itemize}

\end{itemize}

The Inno-EAV framework is proposed to solve problems common to multiple application scenarios, and it is compatible with a multitude of environmental goals and application protocols. Moreover, its architectural and design decisions  does not only manifest but also promote qualities such as modularity, re-usability, modifiability, install-ability, adaptability, authenticity, integrity, interoperability as specified in the ISO/IEC 25010 \cite{ISOIEC2019}.

\section{Design overview}
\label{sec:framework}

 \subsection{Description of Problem}
\label{sec:description}

Traditional Electric charging billing systems requires humans interaction which may be time consuming and exhausting. The Inno-EAV framework will use wireless networks to employ secure machine to machine interaction and perform charging and billing process seamlessly without any manual assistance of humans.
 \subsection{System Operation}
\label{sec:usecase}

The normal flow of Inno-EAV user use case is as follows:

\begin{itemize}
   \item The car owner connects to the same network of the charging station, and send the JSON file to it. In Inno-EAV, we used JSON files as its a light, easy to read way of transmitting data over networks as well as being a native to javascript which makes it easier to extract data.
   \item The charging station verifies the car details by sending the file to the database system  in compliance with authenticity and integrity qualities \cite{ISOIEC2019}.
   \item The charging station then asks for the electricity amount needed.
   \item The car enters the amount.
    \item The bill is sent to the car.
    \item The car pays the bill to the charging station.
\end{itemize}

\begin{figure}[!t]
    \centering
    \includegraphics[width=1.0\textwidth]{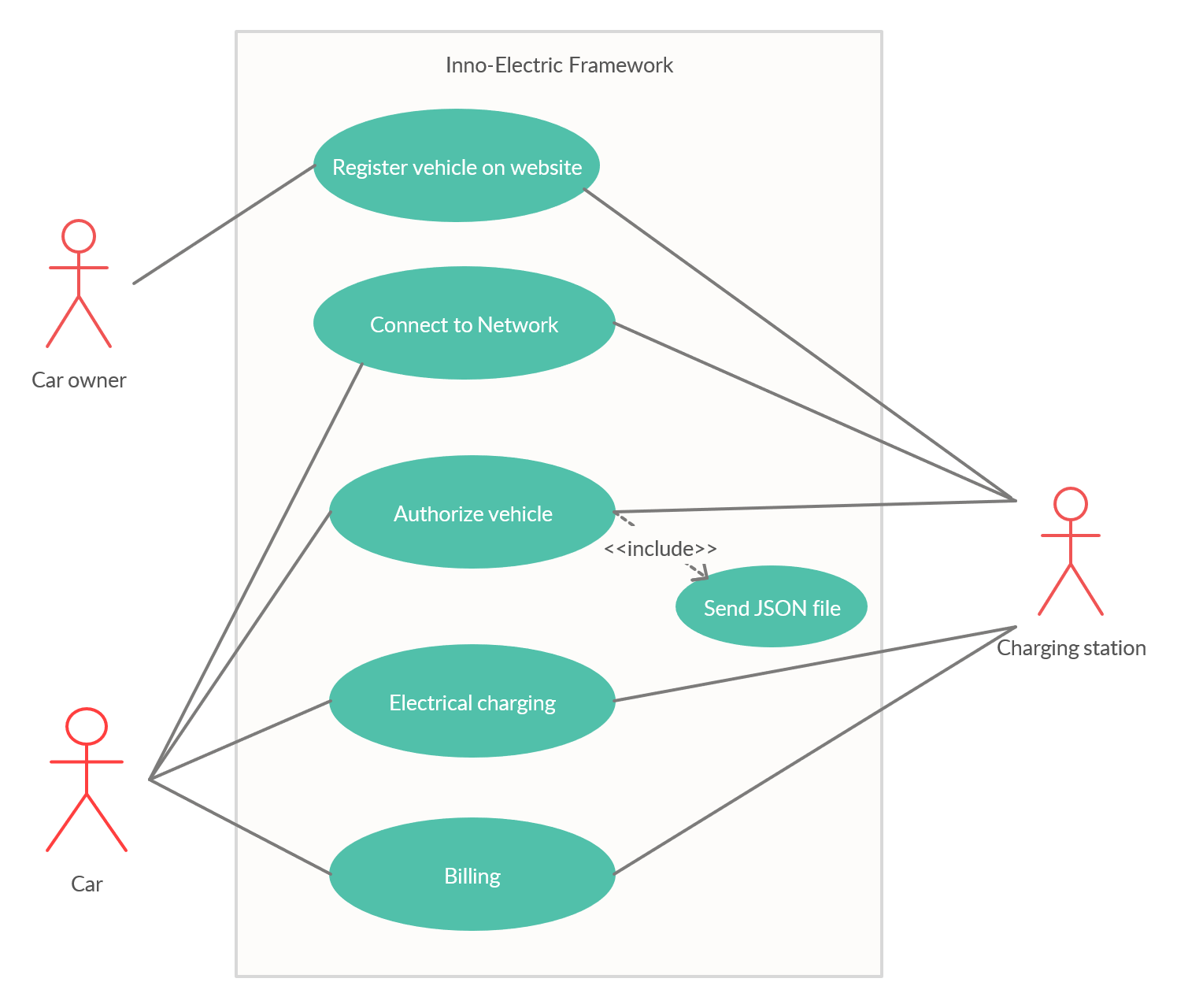}
    \caption{\textbf{UML Use case diagram explanation for Inno-EAV Framework.} Inno-EAV enables the automatic charging of the electric car. The use case begins when an EAV owner connects to the same network as the charging station which it is registered to, and to send the JSON file to the charging station. The system sends the JSON file to the database system for verification. If the database system approves the car details, the server asks for the amount of electric charge and issues a receipt. The transactions are recorded in the database.}
    \label{fig:usecase}
 \end{figure}
 \subsection{Charging Station Registration}
\label{sec:registration}

In M2M communication between server and client, a file is required to be sent to the server at which is often unique for every car. This file has the following attributes: (Name, Email, Phone and Personal-ID) of the car owner and (Model-name, Model-year, and Date purchased) of the car. Each ID is unique, and the combination of the car owner ID and his car is also unique, likewise the ID of charging stations. However, after the file is sent to the server, it checks the combination of the car ID and owner ID against the database for authorization. Charging station registration is therefore essential for assigning the car of each owner to a specific charging station at which it can be charged from in the future. Our team created a web application aiming to ease the registration process for car owners to register their car to a particular station, and thus be able to charge from this station in the future. Therefore, Inno-EAV-Core Charging station registration offers a unique tool to perform this registration quickly and robustly on a wide variety of datasets.
 
The owner will be asked to enter his personal information through form filling, along with his car information,  the information about the charging station where the car is to be registered. In order to complete the authorization step successfully, it is necessary to enter the same details of the vehicle which will charge from the station.
 
\begin{figure}[!t]
    \centering
    \includegraphics[width=1.0\textwidth]{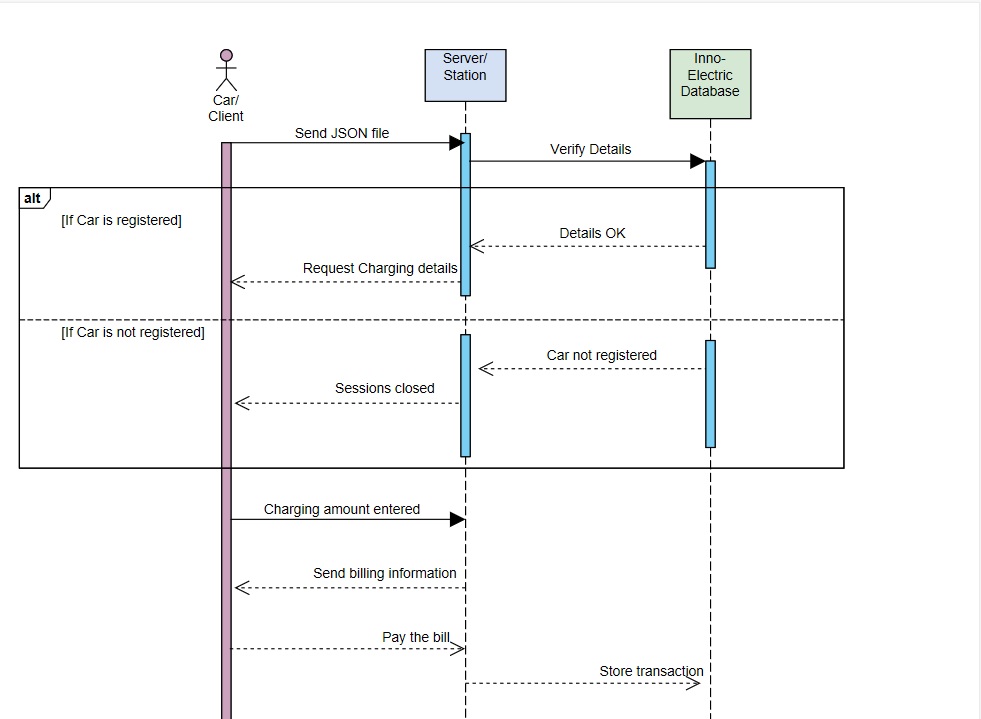}
    \caption{\textbf{Inno-EAV sequence diagram explanation.} Allows any electric car owner to charge their electric car automatically. This sequence begins when an EAV owner connects to the same network as the charging station which it is registered to, it sends the JSON file to the charging station. The system sends the JSON file to the database system for verification. If the database system approves the car details, the server asks for the amount of electric charge and issues a receipt, afterwards, the user pays the bill, and the transaction is stored in the database. Otherwise, the session is closed, and the server waits for another client on the same port.}
    \label{fig:sequence}
 \end{figure}

The sequence diagram begins after the EAV connects to the same WIFI network as the charging station. It has the same steps as refered in section 2, but the additional step happens when the car details are invalid/ car is not registered in the network. In that case, the car is not allowed to charge from this station and the session is closed as shown in Fig \ref{fig:sequence}.

\section{Database System Design}
\label{sec:database}

Inno-EAV-Core performs the charging station registration by creating a new tuple representing each relation entry, where the attributes values for each relation entry corresponds to the value from which the user entered at the registration form. For cases in which these entries violate the key constraints, an error message is used to warn the user, and the registration is not completed successfully. Because the violation of key constraints can happen with different keys in the same entry, the query execution is done after obtaining all the attribute values from the user for each of the car, the owner and the charging station relations, so the violation can also quickly checked and the user asked to correct his entry and enter a valid/unique key.
 
\begin{figure}[!t]
    \centering
    \includegraphics[width=1.0\textwidth]{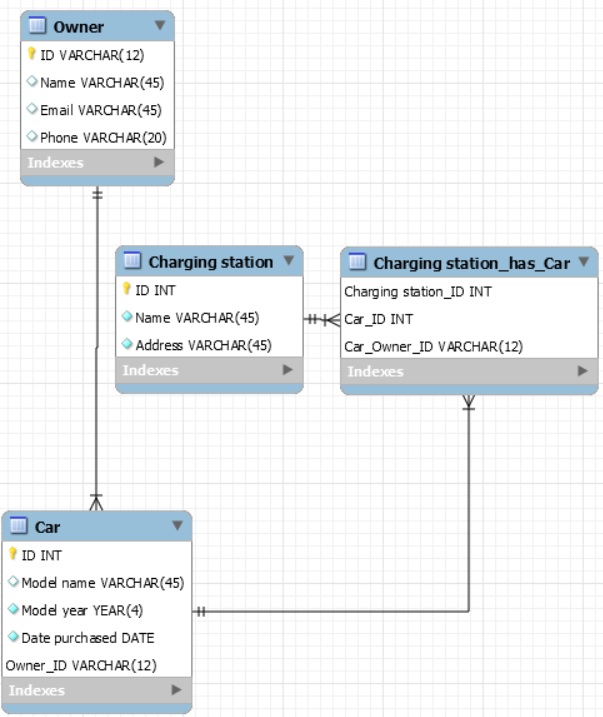}
    \caption{\textbf{Registration Database System Entity Relationship Diagram with Inno-EAV-Core.} \textbf{a)} Owner of Electric-car table representation consisting of four main attributes which are: ID, name, Email and Phone \textbf{b)} Car table representation with four main attributes: ID, Model name, Model year, and Date-purchased and one inherited attribute: Owner-ID \textbf{c)} Charging station table representation with three main attributes: ID, Name, and Address \textbf{d)} Charging station-has-car table representation with three attributes: Charging-station-id, Car-id, Car-owner-ID }
    \label{fig:Relations}
 \end{figure}

To generate these key constraints, Inno-EAV-Core first executes 'Owner' relation query, (Fig. \ref{fig:Relations}). This is achieved by storing the owner's information as a new tuple in the 'Owner' relation, the ID attribute is considered a primary key for the 'Owner' table, as shown in Fig. \ref{fig:Relations}. Then the next step is to execute the 'Car' relation query, which also has a primary key 'ID' representing the car's unique ID  (Fig. \ref{fig:Relations}b). For the given schema, each car entry has a foreign key representing the car's owner. Finally, Inno-EAV-Core can then execute the 'Charging station' query to enter the tuple representing the charging station information that the user can charge from in the future (See Fig. \ref{fig:Relations}).
 
Furthermore, the entries should contain non-repeated key entries uniquely representing each owner, car or charging station. A typical entry for this framework is three tuples with valid key constraints. Following the tuples creation of this entry, Inno-EAV-Core will create another tuple for 'charging station-has-car' relation which represents a many-to-many relationship as described in the following section, This can be used to assign each car to a charging station depending on its unique (Owner-ID, Car-ID) combination. Given that all the entries are valid, therefore, four new tuples will be added to each relation. The relations are: Owner, Car, Charging station, and Charging station-has-car, Inno-EAV-Core can then search the database for the car asking to charge from the station, which is known as authorization step, the car will be accepted or refused depending on whether its registered in the charging station or not (Fig. \ref{fig:Relations}).
 
\subsection{Entity Relationship Diagram}
\label{sec:erm}

As part of the Inno-EAV framework, we include our database entity-relationship diagram (ERD) designed to match our system's needs, which is required to explain the relationships between different relations in the schema.
As shown in Figure \ref{fig:Relations}, the database has been constructed to keep track of the car owners, cars and charging station of a city. A charging station has several cars. It is desired to keep track of the owners owning each vehicle for each charging station, their name, ID, email, and phone the game. Each owner can have more than one car, so the relationship between car and owner relation is one-to-many as can be seen in the diagram. A charging station has more than one car, and a car can be registered in more than one charging station, which yields a many-to-many relationship.

\section{Inno-EAV Implementation}
\label{sec:implementation}

Inno-EAV is installed as a standard set of modules, and it is fully Java-based open-source. Each of its modules possesses its separate manual and documentation. To date, it encompasses two additional Java Archive (JAR) packages (Apache, JSON) alongside with the existing original packages in the Java JDK. The modular nature of Inno-EAV allows its components to be updated independently and the framework to be easily extended by appending new packages hence promoting the maintainability qualities: modularity, re-usability, analysbility, modifiability and test-ability \cite{ISOIEC2019}.

 \subsection{M2M Communication}
\label{sec:network}

\begin{figure}[!t]
\centering
\includegraphics[width=1.0\textwidth]{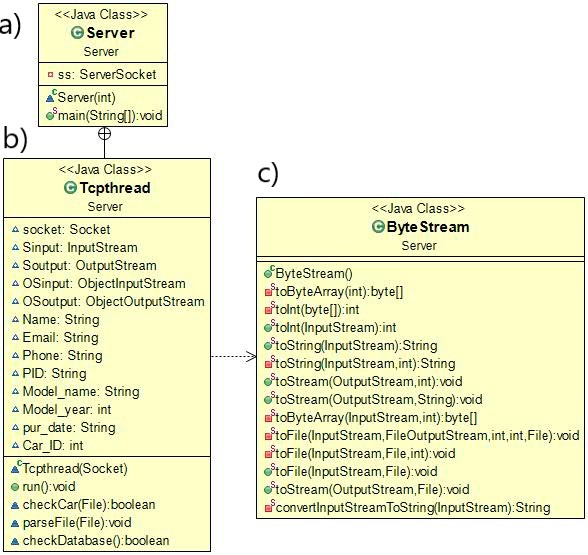}
\caption{\textbf{UML class diagram for Server package.} \textbf{a)} Image of Server class diagram with two main functions: Main, and Server. The latter class is used to specify the port number of connection. \textbf{b)} TCP thread class which extends the Thread class and is run inside Server routine shown in a). \textbf{c)} Byte Stream class used to send and get output streams to and from the client}
\label{fig:ServerPackage}
\end{figure}

M2M communication commonly occurs during the acquisition of the same network to the server and the client, often as a result of the client connecting to the same network where the server is connected. We assume that the network IP address of the charging station is the server with a static IP address. While different clients trying to charge from this server will have a different IP address each time they connect to the wifi network, the charging stations will still be prone to have different IP address each time they connect to the network (Fig. \ref{fig:ServerPackage}). However, in our case where the charging station is considered to be the server, we assume that it uses the same IP address each time it connects to the network and sets up a fixed static IP address for it.
 
Inno-EAV breaks the task of network communication into two distinct parts: Server socket creation from the server, followed by socket creation from the client. \par \textbf{Server - }As a first step, Inno-EAV-Core charging station creates the server socket which represents it and waits for a connection from the client on a specific port number, which is the listening port number (Fig. \ref{fig:ServerPackage}b). The number of the port is previously agreed on by both stakeholders in the connection that is later achieved as the server socket accepts the socket trying to connect to the listening port using socket programming in Java. Once the connection is established, the data of the client can be directly retrieved by getting the input stream using an instance of the ObjectInputStream class shown in (Fig. \ref{fig:ServerPackage}b). The data retrieving process will, however, change the current input stream of the listening port, which can have an influence on further stream writing required to exchange information between the client and the server.
 
For the specific case of our project with Object output stream, there will only be a few writes to the stream of the listening port, as there is only a few information to be exchanged between stakeholders. The first input stream to the server will be the file name of the client, such as 'test.json' for instance. As an option Inno-EAV-Core can store the incoming input lines in the stream within a new file in the sent documents, thus converting the input stream to a new file containing the client's information.
 
 \begin{figure}[!t]
    \centering
    \includegraphics[width=0.6\textwidth]{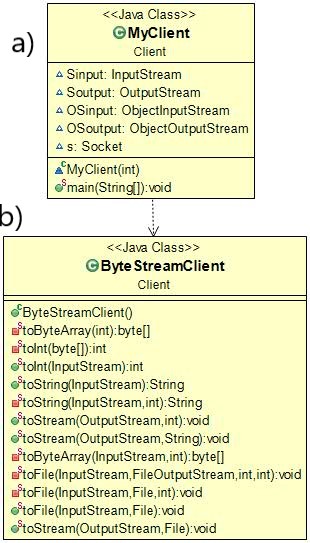}
    \caption{\textbf{UML class diagram for Client package.} \textbf{a)} Image of MyClient class diagram with two main functions: Main, and MyClient. The latter class is used to specify the port number of connection. \textbf{b)} Byte Stream class used to output and get output streams to and from the client}
    \label{fig:ClientPackage}
 \end{figure}

 \textbf{Client - }Client package in Inno-EAV differs slightly from the server package, as it creates the socket with the specific port number and the fixed IP address of the server that we have set in previous steps. This allows the connection to be systematic once the client enters the region of the network to which the server is connected. Furthermore, Inno-EAV-Server (as described in the previous section) can accept the socket creation (Fig.\ref{fig:ClientPackage}a) created by Inno-EAV-Client and use this connection directly to retrieve the data from the client.

As we discussed later, The sequence of Inno-EAV begins when the car connects to the same network that the charging station is connected to, as explained in and (Fig.\ref{fig:ClientPackage}) shows the sequence diagram of Inno-EAV.

\section{Conclusion and Future Work}
\label{sec:conclusion}

In this paper, we investigated the area of Intelligent Vehicle and Transportation Systems proposing a versatile open-source framework developed for improving and automating the charging process of electric vehicles and establishing the foundation of M2M economy using WiFi networks. The networked communication is done without the interference of human, wirelessly through WIFI network. In our proposed scheme, the cars are considered clients and the charging stations are servers. Clients are authenticated with the charging station after they send the file containing the car information to the charging station through a secure TCP connection. Car registration is done using our web application, at which the owners fill in their information and then be assigned to a specific charging station — meanwhile, the electric power service provider bills the vehicles on per charging palate basis. Our proposed scheme provides a secure and privacy-aware bidirectional audit, where the billing process is valid by both parties \cite{innopolis2017}. Moreover, we also present the game-theoretic approach to validate the bidirectional audit.

In the future, we aim to implement the system to have a more in-depth insight into the performance and security issues that can emerge when 
computing has an high level of pervasiveness \cite{DragoniGM16}. Moreover, we also aim to develop the Inno-EAV system to work on a larger scale. To this end, we intend to deploy DevOps practices for faster feature delivery, high quality assurance and product ownership \cite{lwakatare2019devops}. We will also focus on a more robust mechanism where the vehicles will have a choice to buy the charge according to their convenience. We will address the complexity involved in such robust charging and its billing mechanism and can thus be used to optimise acquisition workflows \cite{innopolis2017}. Furthermore, the current framework is fundamentally monolithic-based. To cope with scalability issues, we may in future refine it into a microservice-based one \cite{DragoniLLMMS17} and demonstrate how it can be adopted for projects by DevOps teams.

\bibliographystyle{splncs04}
\bibliography{bibliography}

\begin{thebibliography}{10}
\providecommand{\url}[1]{\texttt{#1}}
\providecommand{\urlprefix}{URL }
\providecommand{\doi}[1]{https://doi.org/#1}

\bibitem{history}
Sit back, relax, and enjoy a ride through the history of self-driving cars.
  \url{https://www.digitaltrends.com/cars/history-of-self-driving-cars-milestones/},
  accessed: 2019-23-8

\bibitem{nhtsa2018reasons}
Critical reasons for crashes investigated in the national motor vehicle crash
  causation survey. Tech. rep., U.S. Department of Transportation, 1200 New
  Jersey Ave, SE, Washington, DC 20590, USA (February 2015),
  \url{https://crashstats.nhtsa.dot.gov/Api/Public/ViewPublication/812115}

\bibitem{afanasyev2019towards}
Afanasyev, I., Kolotov, A., Rezin, R., Danilov, K., Mazzara, M., Chakraborty,
  S., Kashevnik, A., Chechulin, A., Kapitonov, A., Jotsov, V., et~al.: Towards
  blockchain-based multi-agent robotic systems: Analysis, classification and
  applications. arXiv preprint arXiv:1907.07433  (2019)

\bibitem{Bencic18}
Ben{\v{c}}i{\'c}, F.M., {\v{Z}}arko, I.P.: Distributed ledger technology:
  Blockchain compared to directed acyclic graph. In: 2018 IEEE 38th
  International Conference on Distributed Computing Systems (ICDCS). pp.
  1569--1570. IEEE (2018)

\bibitem{Burkhardt18}
{Burkhardt}, D., {Werling}, M., {Lasi}, H.: Distributed ledger. In: 2018 IEEE
  International Conference on Engineering, Technology and Innovation
  (ICE/ITMC). pp.~1--9 (June 2018)

\bibitem{DragoniGM16}
Dragoni, N., Giaretta, A., Mazzara, M.: The internet of hackable things. In:
  Proceedings of 5th International Conference in Software Engineering for
  Defence Applications - {SEDA} 2016, Rome, Italy, May 10th, 2016. pp. 129--140
  (2016)

\bibitem{DragoniLLMMS17}
Dragoni, N., Lanese, I., Larsen, S.T., Mazzara, M., Mustafin, R., Safina, L.:
  Microservices: How to make your application scale. In: Perspectives of System
  Informatics - 11th International Andrei P. Ershov Informatics Conference,
  {PSI} 2017, Moscow, Russia, June 27-29, 2017, Revised Selected Papers. pp.
  95--104 (2017)

\bibitem{InnoEAV2019}
Elbanna, Z.: Innopolis electric autonomous vehicle - inno eav (2019)

\bibitem{hua2018}
Hua, S., Zhou, E., Pi, B., Sun, J., Nomura, Y., Kurihara, H.: Apply blockchain
  technology to electric vehicle battery refueling. In: Proceedings of the 51st
  International Conference on System Sciences (2018)

\bibitem{Hussain2018}
{Hussain}, R., {Lee}, J., {Zeadally}, S.: Autonomous cars: Social and economic
  implications. IT Professional  \textbf{20}(6),  70--77 (Nov 2018)

\bibitem{Hussain2019}
{Hussain}, R., {Zeadally}, S.: Autonomous cars: Research results, issues, and
  future challenges. IEEE Communications Surveys Tutorials  \textbf{21}(2),
  1275--1313 (Secondquarter 2019)

\bibitem{innopolis2017}
Hussain, R., Son, J., Kim, D., Nogueira, M., Oh, H., Tokuta, A.O., Seo6, J.:
  Pbf: A new privacy-aware billing framework for online electric vehicles with
  bidirectional auditability  (2017)

\bibitem{ISOIEC2019}
ISO/IEC: Software product quality.
  https://iso25000.com/index.php/en/iso-25000-standards/iso-25010  (2019)

\bibitem{KYRIAKIDIS2015127}
Kyriakidis, M., Happee, R., de~Winter, J.: Public opinion on automated driving:
  Results of an international questionnaire among 5000 respondents.
  Transportation Research Part F: Traffic Psychology and Behaviour
  \textbf{32},  127 -- 140 (2015)

\bibitem{lwakatare2019devops}
Lwakatare, L.E., Kilamo, T., Karvonen, T., Sauvola, T., Heikkil{\"a}, V.,
  Itkonen, J., Kuvaja, P., Mikkonen, T., Oivo, M., Lassenius, C.: Devops in
  practice: A multiple case study of five companies. Information and Software
  Technology  (2019)

\bibitem{m2mwebsite}
Rouse, M.: What is machine to machine.
  https://internetofthingsagenda.techtarget.com  (2018)

\bibitem{StrugarHMRAL19}
Strugar, D., Hussain, R., Mazzara, M., Rivera, V., Afanasyev, I., Lee, J.: An
  architecture for distributed ledger-based {M2M} auditing for electric
  autonomous vehicles. In: Web, Artificial Intelligence and Network
  Applications - Proceedings of the Workshops of the 33rd International
  Conference on Advanced Information Networking and Applications, {AINA}
  Workshops 2019, Matsue, Japan, March 27-29, 2019. pp. 116--128 (2019)

\bibitem{StrugarHMRLM18}
Strugar, D., Hussain, R., Mazzara, M., Rivera, V., Lee, J.Y., Mustafin, R.: On
  m2m micropayments: a case study of electric autonomous vehicles. In: 2018
  IEEE International Conference on Internet of Things (iThings) and IEEE Green
  Computing and Communications (GreenCom) and IEEE Cyber, Physical and Social
  Computing (CPSCom) and IEEE Smart Data (SmartData). pp. 1697--1700. IEEE
  (2018)

\bibitem{Ting2009}
{Ting}, Z., {Jianjun}, H., {Li}, S., {Jianfeng}, L., {Yan}, M.: Study on
  mobility models in vehicular communication system. In: 2009 2nd IEEE
  International Conference on Broadband Network Multimedia Technology. pp.
  57--61 (Oct 2009). \doi{10.1109/ICBNMT.2009.5347824}

\bibitem{Zhu2019}
{Zhu}, L., {Yu}, F.R., {Wang}, Y., {Ning}, B., {Tang}, T.: Big data analytics
  in intelligent transportation systems: A survey. IEEE Transactions on
  Intelligent Transportation Systems  \textbf{20}(1),  383--398 (Jan 2019).
  \doi{10.1109/TITS.2018.2815678}

\end{thebibliography}
\end{document}